\documentclass[a4paper,DIV15]{scrartcl}

\usepackage{stackengine,scalerel}
\usepackage[english]{babel}
\usepackage{amsmath,amssymb}
\usepackage{pxfonts}
\usepackage{subfigure}
\usepackage{algorithm}
\usepackage[noend]{algpseudocode}

\usepackage{amsmath}
\usepackage{graphicx}
\usepackage{tikz}
\usepackage[strings]{underscore}

\newcommand{\RR}{\mathbb{R}}

\newcommand{\newterm}[1]{\emph{#1}}
\newcommand{\TRUE}{\mathbf{true}}
\newcommand{\FALSE}{\mathbf{false}}
\DeclareMathOperator*{\argmax}{arg\,max}

\newtheorem{definition}{Definition}
\usepackage{url}

\newlength\shlength
\newcommand\xshlongvec[2][0]{\ThisStyle{\setlength\shlength{#1\LMpt}%
  \stackengine{-5.6\LMpt}{$\SavedStyle#2$}{\smash{$\kern\shlength%
    \stackengine{\dimexpr 1.3pt+6.25\LMpt}{$\SavedStyle\mathchar"017E$}%
      {\rule{\widthof{$\SavedStyle#2$}}{\dimexpr.1pt+.5\LMpt}\kern.4\LMpt}{O}{r}{F}{F}{L}\kern-\shlength$}}%
      {O}{c}{F}{T}{S}}}

\author{R\"udiger Ehlers \\ \normalsize University of Bremen and DFKI GmbH, Bremen, Germany}

\title{Formal Verification of Piece-Wise Linear Feed-Forward Neural Networks}

\date{\relax}

\begin{document}
\maketitle

\begin{abstract}
We present an approach for the verification of feed-forward neural networks in which all nodes have a piece-wise linear activation function. Such networks are often used in deep learning and have been shown to be hard to verify for modern satisfiability modulo theory (SMT) and integer linear programming (ILP) solvers.

The starting point of our approach is the addition of a global linear approximation of the overall network behavior to the verification problem that helps with SMT-like reasoning over the network behavior. We present a specialized verification algorithm that employs this approximation in a search process in which it infers additional node phases for the non-linear nodes in the network from partial node phase assignments, similar to unit propagation in classical SAT solving. We also show how to infer additional conflict clauses and safe node fixtures from the results of the analysis steps performed during the search.
The resulting approach is evaluated on collision avoidance and handwritten digit recognition case studies.
\end{abstract}

\section{Introduction}
Many tasks in computing are prohibitively difficult to formalize and thus hard to get right. A classical example is the recognition of digits from images. Formalizing what exactly distinguishes the digit \emph{2} from a \emph{7} is in a way that captures all common handwriting styles is so difficult that this task is normally left to the computer. A classical approach for doing so is to learn a \newterm{feed-forward neural network} from pre-classified example images. Since the advent of \newterm{deep learning} (see, e.g., \cite{DBLP:journals/nn/Schmidhuber15}), the artificial intelligence research community has learned a lot about engineering these networks, such that they nowadays achieve a very good classification precision and outperform human classifiers on some tasks, such as sketch recognition \cite{DBLP:conf/bmvc/YuYSXH15}. Even safety-critical applications such as obstacle detection in self-driving cars nowadays employ neural networks.

But if we do not have formal specifications, how can we assure the safety of such a system? The classical approach to tackle this problem is to construct \emph{safety cases} \cite{Wagner2015}. In such a safety case, we characterize a set of environment conditions under which a certain output is desired and then test if the learned problem model ensures this output under all considered environment conditions. In a self-driving car scenario, we can define an abstract obstacle appearance model all of whose concretizations should be detected as obstacles. Likewise, in a character recognition application, we can define that all images that are \emph{close} to a given example image (by some given metric) should be detected as the correct digit. The verification of safety cases somewhat deviates from the classical aim of formal methods to verify correct system behavior in all cases, but the latter is unrealistic due to the absence of a complete formal specification. Yet, having the means to test neural networks on safety cases would help with certification and also provides valuable feedback to the system engineer.

Verifying formal properties of feed-forward neural networks is a challenging task. Pulina and Tacchella~\cite{DBLP:conf/cav/PulinaT10} present an approach for neurons with non-linear activation functions that only scales to small networks. In their work, they use networks with 6 nodes, which are far too few for most practical applications. They combine counterexample-triggered abstraction-refinement with \newterm{satisfiability modulo theory} (SMT) solving. Scheibler et al.~\cite{DBLP:conf/mbmv/ScheiblerWWB15} consider the bounded model checking problem for an inverse pendulum control scenario with non-linear system dynamics and a non-linear neuron activation function, and despite employing the state-of-the-art SMT solver iSAT3 \cite{DBLP:conf/fmcad/ScheiblerNMFTBF16} and even extending this solver to deal better with the resulting problem instances, their experiments show that the resulting verification problem is already challenging for neural networks with 26 nodes.

In \newterm{deep learning} \cite{DBLP:journals/nn/Schmidhuber15}, many works use networks whose nodes have piece-wise linear activation functions. This choice has the advantage that they are more amenable to formal verification, for example using SMT solvers with the theory of linear real arithmetic, without the need to perform abstract interpretation. 
In such an approach, the solver chooses the \emph{phases} of (some of) the nodes, and then applies a linear-programming-like sub-solver to check if there exist concrete real-valued inputs to the network such that all nodes have the selected phases. The node phases represent which part of the piece-wise linear activation functions are used for each node. It has been observed that the SMT instances stemming from such an encoding are very difficult to solve for modern SMT solvers, as they need to iterate through many such phase combinations before a problem instance is found to be satisfiable or unsatisfiable \cite{DBLP:journals/corr/KatzBDJK17,DBLP:journals/aicom/PulinaT12}. Due to the practical importance of verifying piecewise-linear feed-forward neural networks, this observation asks for a specialized approach for doing so.

Huang et al.~\cite{DBLP:journals/corr/HuangKWW16} describe such an approach that is based on propagating constraints through the layers of a network. The constraints encode regions of the input space of each layer all of whose points lead to the same overall classification in the network. Their approach is partially based on discretization and focusses on robustness testing, i.e., determining the extent to which the input can be altered without changing the classification result. They do not support general verification properties.
Bastiani et al.~\cite{DBLP:conf/nips/BastaniILVNC16} also target robustness testing and define an abstraction-refinement 
constraint solving loop	to test a network's robustness against adversarial pertubations. They also employ the counter-examples that their approach finds to learning more robust networks.
Katz et al.~\cite{DBLP:journals/corr/KatzBDJK17} provide an alternative approach that allows to check the input/output behavior of a neural network with linear and so-called \newterm{ReLU} nodes against convex specifications. Many modern network architectures employ these nodes. They present a modification of the \newterm{simplex algorithm} for solving linear programs that can also deal with the constraints imposed by ReLU nodes, and they show that their approach scales orders of magnitudes better than when applying the SMT solvers \texttt{MathSAT} or \texttt{Yices} on SMT instances generated from the verification problems.

Modern neural network architectures, especially those for image recognition, however often employ another type of neural network node that the approach by Katz et al.~does not support: \newterm{MaxPool} nodes. They are used to determine the strongest signal from their input neurons, and they are crucial for \newterm{feature detection} in complex machine learning tasks. In order to support the verification of safety cases for machine learning applications that make use of this node type, it is thus important to have verification approaches that can efficiently operate on networks that have such nodes, without the need to simulate MaxPool nodes by encoding their behavior into a much larger number of ReLU nodes.

In this paper, we present an approach to verify neural networks with piece-wise linear activation functions against convex specifications. The approach supports all node types used in modern network network architectures that only employ piece-wise linear activation functions (such as MaxPool and ReLU nodes). The approach is based on combining satisfiability (SAT) solving and linear programming and employs a novel linear approximation of the overall network behavior. This approximation allows the approach to quickly rule out large search space parts for the node phases from being considered during the verification process. While the approximation can also be used as additional constraints in SMT solving and improves the computation times of the SMT solver, we apply it in a customized solver that uses the \newterm{elastic filtering} algorithm from \cite{DBLP:journals/informs/ChinneckD91} for minimal infeasible linear constraint set finding in case of conflicts, and combine it with a specialized procedure for inferring implied node phases. Together, these components lead to much shorter verification times. We apply the approach on two cases studies, namely collision avoidance and character recognition, and report on experimental results.
We also provide the resulting solver and the complete tool-chain to generate verifiable models using the Deep Learning framework \texttt{Caffe}~\cite{jia2014caffe} as open-source software.

\section{Preliminaries}

\paragraph{Feed-Forward Neural Networks:} We consider \newterm{multi-layer} (\newterm{Perceptron}) networks with \newterm{linear}, \newterm{ReLU}, and \newterm{MaxPool} nodes in this paper. Such networks are formally defined as directed acyclic weighted graphs $G = (V,E,W,B,T)$, where $V$ is a set of nodes, $E \subset V \times V$  is a set of edges, $W : E \rightarrow \RR$ assigns a weight to each edge of the network, $B : V \rightarrow \RR$ assigns a \newterm{node bias} to each node, and $T$ assigns a \emph{type} to each node in the network from a set of available types $\mathcal{T} \in \{\mathit{input}, \mathit{linear}, \mathit{ReLU}, \mathit{MaxPool}\}$. Nodes without incoming edges are called \newterm{input nodes}, and we assume that $T(v) = \mathit{input}$ for every such node $v$. Vertices that have no outgoing edge are also called \newterm{output nodes}.

A feed-forward neural network with $n$ input nodes and $m$ output nodes represents a function $f : \RR^n \rightarrow \RR^m$. Given assignments $\mathit{in} : \{1, \ldots, n\} \rightarrow V$ and  $\mathit{out} : \{1, \ldots, m\} \rightarrow V$ that define the orders of the input and output nodes (so that we can feed elements from $\RR^n$ to the network to obtain an output from $\RR^m$), and some input vector $(x_1, \ldots, x_n) \in \RR^n$, we can define the network's behavior by a node value assignment function $a:V \rightarrow \RR$ that is defined as follows:
\begin{itemize}
\item For every node $v$ with $T(v) = \mathit{input}$, we set $a(v) = x_j$ for $j={\mathit{in}^{-1}(v)}$,
\item For every node $v$ with $T(v) = \mathit{linear}$, we set $a(v) = \sum_{v' \in V, (v',v) \in E} W((v',v)) \cdot a(v') + B(v)$.
\item For every node $v$ with $T(v) = \mathit{ReLU}$, we set $a(v)= \max(B(v)+\sum_{v' \in V, (v',v) \in E} \allowbreak{} W((v',v)) \cdot a(v'),0)$.
\item For every node $v$ with $T(v) = \mathit{MaxPool}$, we set $a(v) = \max_{v' \in V, (v',v) \in E} a(v')$.
\end{itemize}
Function $f$'s output for $(x_1,\ldots,x_n)$ is defined to be $(a(\mathit{out}(1)), \allowbreak{}\ldots, \allowbreak{} a(\mathit{out}(m)))$. Note that the weights of the edges leading to $\mathit{MaxPool}$ nodes and their bias values are not used in the definition above. 
Given a node value assignment function $a:V \rightarrow \RR$, we also simply call $a(v)$ the \newterm{value} of $v$. If for a ReLU node $v$, we have $s(v)<0$ for $s(v) = B(v)+\sum_{v' \in V, (v',v) \in E} \allowbreak{} W((v,v')) \cdot a(v')$, and hence $a(v) = 0$, we say that node $n$ is in the $\leq 0$ phase, and for $s(v) \geq 0$, and hence $a(v) \geq 0$, we say that it is in the $\geq 0$ phase. If we have $s(v)=0$, then it can be in either phase. For a $\mathit{MaxPool}$ node $v$, we define it to be in phase $e \in E \cap (V \times \{v\})$ if $a(v)=a(v')$ for $e = (v',v)$. If multiple nodes with edges to $v$ have the same values, then node $v$ can have any of the respective phases.

Modern neural network architectures are \newterm{layered}, i.e., we have that every path from an input node to an output node has the same length. For the verification techniques given in this paper, it does however not matter whether the network is layered.
Networks can also be used to \newterm{classify} inputs. In such a case, the network represents a function $f' : \RR^n \rightarrow \{1, \ldots, m\}$ (for some numbering of the classes), and we define $f'(x_1, \ldots, x_n) = \argmax_{i \in \{1, \ldots, m\}} y_i$ for $(y_1, \ldots, y_m) = f(x_1, \ldots, x_n)$.

We do not discuss here how neural networks are learned, but assume networks to be given with all their edge weights and node bias values. Frameworks such as \texttt{Caffe}~\cite{jia2014caffe} provide ready-to-use functionality for learning edge weights and bias values from databases of examples, i.e., tuples $(x_1, \ldots, x_n, y_1, \ldots, y_m)$ such that we want the network to induce a function $f$ with $(x_1, \ldots, x_n) = (y_1, \ldots, y_m)$. Likewise, for classification problems, the databases consist of tuples $(x_1, \ldots, x_n, c)$ such that we want the network to induce a function $f'$ with $f'(x_1, \ldots, x_n) = c$.
When using a neural network learning tool, the architecture of the network, i.e., everything except for the weights and the node bias values, is defined up-front, and the framework automatically derives suitable edge weights and node bias values. There are other node types (such as \newterm{Softmax} nodes) that are often used during the learning process, but removed before the deployment of the trained network, and hence do not need to be considered in this work. Also, there are network layer types such as \newterm{convolutional layers} that have special structures. From a verification point of view, these are however just sets of linear nodes whose edges share some weights, and thus do not have to be treated differently.

\paragraph{Satisfiability Solvers:}
Satisfiability (SAT) solvers check if a Boolean formula has a satisfying assignment. 
The formula is normally required to be in conjunctive normal form, and thus consists of \newterm{clauses} that are connected by \newterm{conjunction}. Every clause is a disjuction of one of more \newterm{literals}, which are Boolean variables or their negation.
A SAT solver operates by successively building a valuation of the Boolean variables and \newterm{backtracking} whenever a conflict of the current \newterm{partial valuation} and a clause has been found. To achieve a better performance, SAT solvers furthermore perform \newterm{unit propagation}, where the partial assignment is extended by literals that are the only remaining ones not yet violated by the partial valuation in some clause. Additionally, modern solvers perform \newterm{clause learning}, where clauses that are implied by the conjunction of some other clauses are lazily inferred during the search process, and select variables to branch on using a \newterm{branching heuristic}. Most modern solvers also perform \newterm{random restarts}. For more details on SAT solving, the interested reader is referred to \cite{FM09HBSAT}.

\paragraph{Linear Programming:} Given a set of linear inequalities over real-valued variables and a linear optimization function (which together are called a \newterm{linear program}), the linear programming problem is to find an assignment to the variables that minimizes the objective function and fulfills all constraints. Even though linear programming was shown to have polynomial-time complexity, it has been observed that in practice \cite{Kroening2008}, it is often faster to apply the \newterm{Simplex algorithm}, which is an exponential-time algorithm.

\paragraph{Satisfiability Modulo Theory Solving:}
SAT solvers only support Boolean variables. For problems that can be naturally represented as a Boolean combination of constraints over other variable types, Satisfiability Modulo Theory (SMT) solvers are normally applied instead. An SMT solver combines a SAT solver with specialized decision procedures for other theories (such as, e.g., the theory of linear arithmetic over real numbers).

\section{Efficient Verification of Feed-forward Neural Networks}
In this paper, we deal with the following verification problem:
\begin{definition}
\label{def:mainProblem}
Given a feed-forward neural network $G$ that implements a function $f : \RR^n \rightarrow \RR^m$, and a set of linear constraints $\psi$ over the real-valued variables $V = \{x_1, \ldots, x_n, y_1, \ldots, y_m \}$, the neural net (NN) verification problem is to find a node value assignment function $a$ for $V$ that fulfils $\psi$ over the input and output nodes of $G$ and for which we have $f(x_1, \ldots, x_n) = (y_1, \ldots, y_m)$, or to conclude that no such node value assignment function exists.
\end{definition}
The restriction to conjunctions of linear properties in Definition~\ref{def:mainProblem} was done for simplicity. Verifying arbitrary Boolean combinations of linear properties can be fitted into Definition~\ref{def:mainProblem} by encoding them into the structure of the network itself, so that an additional output neuron $y_{\mathit{add}}$ outputs a value $\geq 0$ if and only if the property is fulfilled. In this case, $\psi$ is then simply $y_{\mathit{add}} \geq 0$.

There are multiple ways to solve the neural network (NN) verification problem. The encoding of an NN verification problem to an SMT problem instance is straight-forward, but yields instances that are difficult to solve even for modern SMT solvers (as the experiments reported on in Section~\ref{sec:experiments} show). 
As an alternative, we present a new approach that combines 1) linear approximation of the overall NN behavior, 2) irreducible infeasible subset analysis for linear constraints based on elastic filtering \cite{DBLP:journals/informs/ChinneckD91}, 3) inferring possible safe node phase choices from feasibility checking of partial node phase valuations, and 4) performing unit-propagation-like reasoning on node phases. 
We describe these ideas in this section, and present experimental results on a tool implementing them in the next section.

Starting point is the combination of a linear programming solver and a satisfiability solver. 
We let the satisfiability solver guide the search process. It determines the phases of the nodes and maintains a set of constraints over node phase combinations. On a technical level, we allocate the SAT variables $x_{(v,\leq 0)}$ and $x_{(v,\geq 0)}$ for every ReLU node $v$, and also reserve variables $x_{(v,e)}$ for every MaxPool node $v$ and every edge $e$ ending in $v$. The SAT solver performs unit propagation, clause learning, branching, and backtracking as usual, but whenever the solver is about to branch, we employ a linear programming solver to check a linear approximation of the network behavior (under the node phases already fixed) for feasibility. Whenever a conflict is detected, the SAT solver can then learn a conflict clause. Additionally, we infer implied node phases in the search process.

We describe the components of our approach in this section, and show how they are combined at the end of it.

\subsection{Linear Approximation of Neural Network Value Assignment Functions}
\label{subsec:linearApprox}

Let $G = (V,E,W,B,T)$ be a network representing a function $f : \RR^n \rightarrow \RR^m$. We want to build a system of linear constraints using $V$ as the set of variables that closely approximates $f$, i.e., such that every node value assignment function $a$ is a correct solution to the linear constraint system, and the constraints are as tight as possible. The main difficulty in building such a constraint system is that the $\mathit{ReLU}$ and $\mathit{MaxPool}$ nodes do not have linear input-output behavior (until their phases are fixed), so we have to approximate them linearly.

\begin{figure}[tb]
\centering\begin{tikzpicture}
\draw[dashed] (-1.5,0) -- (2,2);
\path[fill=black!20!white] (-1.5,0) -- (0,0) -- (2,2) -- cycle;

\draw[->] (-3,0) -- (3,0) node[right] {$c$};
\draw[->] (0,-0.5) -- (0,2) node[right] {$\,d$};

\draw[very thick] (-3,0) -- (0,0) -- (2,2);

\draw (2.0,0.2) -- (2.0,-0.2) node[below] {$u$};
\draw (-1.5,0.2) -- (-1.5,-0.2) node[below] {$l$};

\end{tikzpicture}
\caption{The activation function of a $\mathit{ReLU}$ node, with a linear over-approximation drawn as filled area.}
\label{fig:ReLU}
\end{figure}
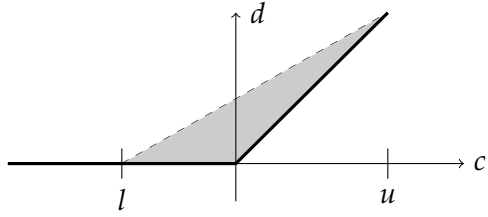

Figure~\ref{fig:ReLU} shows the activation function of a $\mathit{ReLU}$ node, where we denote the weight\-ed sum of the input signals to the node (and its bias) as variable $c$. The output of the node is denoted using the variable $d$.
If we have upper and lower bounds $[l,u]$ of $c$, then we can approximate the relationship between $c$ and $d$ by the constraints $d \geq 0$, $d \geq c$, and $d \leq \frac{u \cdot (c - l)}{u-l}$, all of which are linear equations for constant $u$ and $l$. This yields the set of allowed value combinations for $c$ and $d$ drawn as the filled area in Figure~\ref{fig:ReLU}.

Obviously, this approach requires that we know upper and lower bounds on $c$. However, even though neural networks are defined as functions from $\RR^n$, bounds on the input values are typically known. For example, in image processing networks, we know that the input neurons receive values from the range $[0,1]$. In other networks, it is common to \newterm{normalize} the input values before learning the network, i.e., to scale them to the same interval or to $[-1,1]$. This allows us to use classical \emph{interval arithmetic} on the network to obtain basic lower and upper bounds $[l,u]$ on every node's values.

For the case of $\mathit{MaxPool}$ nodes, we can approximate the behavior of the nodes linearly similarly to the ReLU case, except that we do not need upper bounds for the nodes' values. Let $c_1, \ldots, c_k$ be the values of nodes with edges leading to the $\mathit{MaxPool}$ node, $l_1, \ldots, k_k$ be their lower bounds, and $d$ be the output value of the node. We instantiate the following linear constraints:
\begin{equation*}
\bigwedge_{i \in \{1, \ldots, k\}} (d \geq c_i) \ \ \wedge \ \ (c_1 + \ldots + c_k \geq d + \sum_{i \in \{1,\ldots,k\}} l_i - \max_{i \in \{1,\ldots,k\}} l_i)
\end{equation*}
Note that these are the tightest linear constraints that can be given for the relationship between the values of the predecessor nodes of a $\mathit{MaxPool}$ node and the node value of the $\mathit{MaxPool}$ node itself.

After a linear program that approximates the behavior of the overall network has been built, we can use it to make all future approximations even tighter. To achieve this, we add the problem specification $\psi$ as constraints and solve, for every variable $v \in V$, the resulting linear program while minimizing first for the objective functions $1 \cdot v$, and then doing the same for the objective function $-1 \cdot v$. This yields new tighter lower and upper bounds $[l,u]$ for each node (if the network has any ReLU nodes), which can be used to obtain a tighter linear program.
Including the specification in the process allows us to derive tighter bounds than we would have found without the specification. The whole process can be repeated several times: whenever new upper and lower bounds have been obtained, they can be used to build a tighter linear network approximation, which in turn allows to obtain new tighter upper and lower bounds.

\subsection{Search process and Infeasible Subset Finding}
\label{subsec:basicSearchProcess}

Given a phase fixture for all ReLU and MaxPool nodes in a network, checking if there exists a node value assignment function with these phases (and such that the verification constraint $\psi$ is fulfilled) can be reduced to a linear programming problem. For this, we extend the linear program built with the approach from the previous subsection (with $V$ as the variable set for the node values) by the following constraints:
\begin{itemize}
\item For every $\leq 0$ phase selected for a ReLU node $v$, we add the constraints $v = 0$ and $\sum_{(v',v) \in E} W((v', \allowbreak{} v)) \cdot v' + B(v) \leq 0$.
\item For every $\geq 0$ phase selected for a ReLU node $v$, we add the constraint $v \geq \sum_{(v',v) \in E} \allowbreak{} W((v',v)) \cdot v' + B(v)$.
\item For every phase $(v',v)$ selected for a MaxPool node $v$, we add the constraint $v = v'$.
\end{itemize}
If we only have a partial node phase selection, we add these constraints only for the fixed nodes.
If the resulting linear program is infeasible, then we can discard all extensions to the partial valuation from consideration in the search process. This is done by adding a \newterm{conflict} clause that rules out the Boolean encoding of this partial node phase selection, so that even after restarts of the solver, the reason for infeasibility is retained.

However, the reasons for conflicts often involve relatively few nodes, so shorter conflict clauses can also be learned instead (which makes the search process more efficient). To achieve this, we employ \emph{elastic filtering} \cite{DBLP:journals/informs/ChinneckD91}. In this approach, all of the constraints added due to node phase selection are weakened by \emph{slack variables}, where there is one slack variable for each node. So, for example a constraint $\sum_{(v',v) \in E} W((v',v)) \cdot v' + B(v) \leq 0$ becomes $\sum_{(v',v) \in E} W((v',v)) \cdot v' + B(v) - s_v \leq 0$. When running the linear programming solver again with the task of minimizing a weighted sum of the slack variables, we get a ranking of the nodes by how much they contributed to the conflict, where some of them did not contribute at all (since their slack variable had a $0$ value). We then fix the slack variable with the highest value to be $0$, hence making the corresponding constraints strict, and repeat the search process until the resulting LP instance becomes infeasible. We then know that the node phase fixtures that were made strict during this process are together already infeasible, and build conflict clauses that only contain them. We observed that these conflict clauses are much shorter than without applying elastic filtering.

Satisfiability modulo theory solvers typically employ cheaper procedures to compute \newterm{minimal infeasible subsets} of linear constraints, such as the one by Duterte and de Moura~\cite{DBLP:conf/cav/DutertreM06}, but the high number of constraints in the linear approximation of the network behavior that are independent of node phase selections seems to make the approach less well-suited, as our experiments with the SMT solver \texttt{Yices} that uses this approach suggest.

\subsection{Implied Node Phase Inference during Partial Phase Fixture Checking}
\label{subsec:inferredNodeDetection}
In the partial node fixture feasibility checking step from Section~\ref{subsec:basicSearchProcess}, we employ a linear programming solver. However, except for the elastic filtering step, we did not employ an optimization function yet, as it was not needed for checking the feasibility of a partial node fixture.

For the common case that the partial node fixture \emph{is} feasible (in the linear approximation), we define an optimization function that allows us to infer additional infeasible \emph{and} feasible partial node fixtures when checking some other partial node fixture for feasibility.
The feasible fixtures are cached so that if it or a partial fixture of it is later evaluated, no linear programming has to be performed.
Given a partial node fixture to the nodes $V' \subset V$, we use $-1 \cdot \sum_{v \in V \setminus V', T(v) = \mathit{ReLU}} v - \frac{1}{10} \sum_{v \in V \setminus V', T(v) = \mathit{MaxPool}} v$ as optimization function. This choice asks the linear programming solver to minimize the error for the ReLU nodes, i.e, the difference between $a(v)$ and $\max(\sum_{v' \in V, (v',v) \in E} W((v',v)) \cdot a(v')+B(v),0)$ for every assignment $a$ computed in the linear approximation of the network behavior and every ReLU-node $v$. While this choice only minimizes an approximation of the error sum of the nodes and thus does not guarantee that the resulting variable valuation denotes a valid node value assignment function, it often yields assignments in which a substantial number of nodes $v$ \emph{have} a tight value, i.e., have $a(v) = \max(\sum_{v' \in V, (v',v) \in E} W((v',v)) \cdot a(v')+B(v),0)$. 

If $\mathsf{tight}$ is the set of nodes with tight values, $p$ is the partial SAT solver variable valuation that encodes the phase fixtures for the nodes $V'$, and if $p'$ is the (partial) valuation of the SAT variables that encodes the phases of the tight nodes, we can then cache that $p \cup p'$ is a partial assignment that is feasible in the linear approximation. So when the SAT solver adds literals from $p'$ to the partial valuation, there is no need to let the linear programming solver run again.

At the same time, the valuation $a$ (in the linear approximation) can be used to derive an additional clause for the SAT solver. Let $\mathsf{unfixed}$ be the ReLU nodes whose values are not fixed by $p$. If for any node $v \in \mathsf{unfixed}$, we have $a(v)>0$, then we know by the choice of optimization function and the fact that we performed the analysis in a linear approximation of the network behavior, that some node in $v$ needs to be in the $\geq 0$ phase (under the partial valuation $p$). Thus, we can learn the additional clause $\left(\bigvee_{l \in p} \neg l\right) \vee \bigvee_{v \in V, T(v) = \mathit{ReLU}, ((v,\leq 0) \mapsto \TRUE) \notin p} (v,$ $\geq 0)$ for the SAT solver, provided that the values of the MaxPool nodes are valid, i.e, for all MaxPool nodes $v$ we have $a(v)=a(v')$ for some $(v,v') \in E$. This last restriction is why we also included the MaxPool nodes in the optimization function above (but with lower weight).

\subsection{Detecting Implied Phases}
\label{subsec:detectingImpliedPhases}

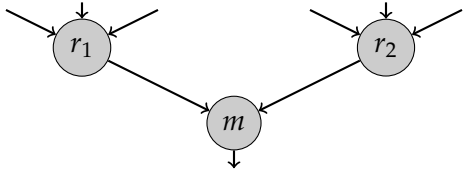
\begin{figure}[bt]
\centering\begin{tikzpicture}
\node[draw,shape=circle,fill=black!20!white] (relu1) at (0,0) {$r_1$};
\node[draw,shape=circle,fill=black!20!white] (relu2) at (4,0) {$r_2$};
\node[draw,shape=circle,fill=black!20!white] (maxpool) at (2,-1) {$m$};
\draw[->,thick] (-1,0.5) -- (relu1);
\draw[->,thick] (-0,0.6) -- (relu1);
\draw[->,thick] (1,0.5) -- (relu1);
\draw[->,thick] (3,0.5) -- (relu2);
\draw[->,thick] (4,0.6) -- (relu2);
\draw[->,thick] (5,0.5) -- (relu2);
\draw[->,thick] (relu1) -- (maxpool);
\draw[->,thick] (relu2) -- (maxpool);
\draw[->,thick] (maxpool) -- +(0,-0.6);
\end{tikzpicture}
\caption{An example neural network part, used in Subsection~\ref{subsec:detectingImpliedPhases}.}
\label{fig:newExcerpt}
\end{figure}

Whenever the SAT solver has fixed a new node phase, the selected phases together may imply other node phases. Take for example the net excerpt from Figure~\ref{fig:newExcerpt}. There are two ReLU nodes, named $r_1$ and $r_2$, and one MaxPool node. Assume that during the initial analysis of the network (Section~\ref{subsec:linearApprox}), it has been determined that the value of node $r_1$ is between $0.0$ and $1.5$, and the value of node $r_2$ is between $0.1$ and $2.0$. First of all, the SAT solver can unconditionally detect that node $r_2$ is in the $\geq 0$ phase. Then, if at some point, the SAT solver decides that node $r_1$ should be in the $\leq 0$ phase, this fixes the value of $r_1$ to $0$. Since the flow out of $r_2$ has a lower bound $>0$, we can then deduce that $m$'s phase should be set to $(r_2,m)$. 

Similar reasoning can also be performed for flow leading into a node. If we assume that the analysis of the initial linear approximation of the network's node functions yields that the outgoing flow of $m$ needs to be between $0.5$ and $0.7$, and the phase of $r_1$ is chosen to be $\leq 0$, then this implies that the phase of $r_2$ must be $\geq 0$, as otherwise $m$ would be unable to supply a flow of $>0$. 

Both cases can be detected without analyzing the linear approximation of the network. Rather, we can just propagate the lower and upper bounds on the nodes' outgoing flows through the network and detect implied phases. Doing so takes time linear in the size of the network, which is considerably faster than making an LP solver call. This allows the detection of implied phases to be applied in a way similar to classical unit propagation in SAT solving: whenever a decision has been made by the solver, we run implied phase detection to extend the partial valuation of the SAT solver by implied choices (which allows to make the linear approximation tighter for the following partial node fixture feasibility checks).

\subsection{Overview of the Integrated Solver}
To conclude this section, let us discuss how the techniques presented in it are combined. Algorithm~\ref{algo:nnVerifier} shows the overall approach. In the first step, upper and lower bounds for all nodes' values are computed. The solver then prepares an empty partial valuation to the SAT variables and an empty list $\mathit{extra}$ in which additional clauses generated by the LP instance analysis steps proposed in this section are stored. The SAT instance is initialized with clauses that enforce that every ReLU node and every MaxPool node has exactly one phase selected (using a \newterm{one-hot encoding}).

In the main loop of the algorithm, the first step is to perform most steps of SAT solving, such as unit propagation, conflict detection \& analysis, and others. We assume that the partial valuation is always labelled by \newterm{decision levels} so that backtracking can also be performed whenever needed. Furthermore, additional clauses from $\mathit{extra}$ are mixed to the SAT instance $\psi$. This is done on a step-by-step basis, as the additional clauses may trigger unit propagation and even conflicts, which need to be dealt with eagerly. After all clauses from $\mathit{extra}$ have been mixed into $\psi$, and possibly the partial valuation $p$ has been extended by implied literals, in line \ref{line:infer}, the approach presented in Sect.~\ref{subsec:detectingImpliedPhases} is applied. If it returns new implied literals (in the form of additional clauses), they are taken care of by the SAT solving steps in line~\ref{line:satSolvingSteps} next. This is because the clauses already in $\psi$ may lead to unit propagation on the newly inferred literals, which makes sense to check as every additional literal makes the linear approximation of the network behavior tighter (and can lead to additional implied literals being detected). Only when all node phases have been inferred, $p$ is checked for feasibility in the linear approximation (line~\ref{line:check}). 

There are two different outcomes of this check: if the LP instance is infeasible, a new conflict clause is generated, and hence the condition in line~\ref{ref:ifClauseLabel} is not satisfied. The algorithm then continues in line~\ref{line:satSolvingSteps} in this case. Otherwise, the branching step of the SAT solver is executed. If $p$ is already a complete valuation, we know at this point that the instance is \emph{satisfiable}, as then the $\mathrm{CheckForFeasibility}$ function just executed operated on an LP problem that is not approximate, but rather captures the precise behavior of the network. Otherwise, $p$ is extended by a decision to set a variable $b$ to $\TRUE$ (for some variable chosen by the SAT solver's variable selection heuristics). Whenever this happens, we employ a plain SAT solver for checking if the partial valuation can be extended to one that satisfies $\psi$. This not being the case may not be detected by unit propagation in line~\ref{line:satSolvingSteps} and hence it makes sense to do an eager SAT check. In case of conflict, the choice of $b$'s value is inverted, and in any case, the algorithm continues with the search.

\algrenewcommand\algorithmicindent{2.3em}

\begin{algorithm}[tb]
\begin{algorithmic}[1]
\Function{VerifyNN}{$V,E,T,B,W$}
\State $(\overrightarrow{\mathit{min}},\overrightarrow{\mathit{max}}) \gets \mathrm{ComputeInitialBounds}(V,E,T,B,W)$ \Comment{Section~\ref{subsec:linearApprox}}
\State $(\overrightarrow{\mathit{min}},\overrightarrow{\mathit{max}}) \gets \mathrm{RefineBounds}(V,E,T,B,W,\overrightarrow{\mathit{min}},\overrightarrow{\mathit{max}})$ \Comment{Section~\ref{subsec:linearApprox}}
\State $p \gets \emptyset$, $\mathit{extra} \gets \emptyset$
\State $\psi \gets \bigwedge_{v \in V, T(v)=\mathit{MaxPool}} \bigvee_{v' \in V, (v',v) \in E} x_{v,(v',v)}$
\State $\psi \gets \psi \wedge \bigwedge_{v \in V, T(v)=\mathit{MaxPool}, v',v'' \in V, v' \neq v'', (v',v) \in E, (v'',v) \in E} (\neg x_{v,(v',v)} \vee \neg x_{v,(v'',v)})$
\State $\psi \gets \psi \wedge \bigwedge_{v \in V, T(v)=\mathit{ReLU}} (x_{v,\leq 0} \vee x_{v,\geq 0}) \wedge (\neg x_{v,\leq 0} \vee \neg x_{v,\geq 0})$ 
\While{$\psi$ has a satisfying assignment}
	\While{$\mathit{extra}$ is non-empty}
		\State Perform unit propagation, conflict detection, backtracking, and clause 
		\State learning for $p$ on $\psi$, while moving the clauses from $\mathit{extra}$ to $\psi$ one-by-one.\label{line:satSolvingSteps}
	\EndWhile
	\State $\mathit{extra} \gets \mathrm{InferNodePhases}(V,E,T,B,W,p,\overrightarrow{\mathit{min}},\overrightarrow{\mathit{max}})$ \Comment{Section~\ref{subsec:detectingImpliedPhases}} \label{line:infer}
	\If{$\mathit{extra}=\emptyset$}
		\State $\mathit{extra} \gets \mathrm{CheckForFeasibility}(V,E,T,B,W,p,\overrightarrow{\mathit{min}},\overrightarrow{\mathit{max}})$ \Comment{Section~\ref{subsec:basicSearchProcess}-\ref{subsec:inferredNodeDetection}} \label{line:check}
		\If{$p \models c$ for all clauses $c \in \mathit{extra}$} \label{ref:ifClauseLabel}
			\If{$p$ is a complete assignment to all variables}
				\State \Return \textbf{Satisfiable}
			\EndIf
			\State Add a new variable assignment $b \!\mapsto \TRUE$ to $p$ for some variable $b$ in $\psi$.
			\If{$p$ cannot be extended to a satisfying valuation to $\psi$}
				\State $p = p \setminus \{b \mapsto \TRUE \} \cup \{b \mapsto \FALSE\}$
			\EndIf
		\EndIf
	\EndIf
\EndWhile
\State \Return \textbf{Unsatisfiable}
\EndFunction
\end{algorithmic}
\caption{Top-level view onto the neural network verification algorithm.}
\label{algo:nnVerifier}
\end{algorithm} 

\section{Experiments}
\label{sec:experiments}
We implemented the approach presented in the preceding section in a tool called \texttt{Planet}. It is written in C++ and bases on the linear programming toolkit \texttt{GLPK} 4.61\footnote{GNU Linear Programming Kit, \url{http://www.gnu.org/software/glpk/glpk.html}} and the SAT solver \texttt{Minisat 2.2.0} \cite{DBLP:conf/sat/EenS03}. While we use \texttt{GLPK} as it is, we modified the main search procedure of \texttt{Minisat} to implement Algorithm~\ref{algo:nnVerifier}. 
We repeat the initial approximation tightening process from Section~\ref{subsec:linearApprox} until the cumulative changes in $\overrightarrow{\mathit{min}}$ and $\overrightarrow{\mathit{max}}$ fall below $1.0$. We also abort the process if 5000 node approximation updates have been performed (to not spend too much time in the process for very large nets), provided that for every node, its bounds have been updated at least three times. 

All numerical computations are performed with \texttt{double} precision, and we did not use any compensation for numerical imprecision in the code apart from using a fixed safety margin $\epsilon = 0.0001$ for detecting node assignment values $a(v)$ to be greater or smaller than other node assignment values $a(v')$, i.e., we actually check if $a(v) \leq a(v')-\epsilon$ to conclude $a(v) \leq a(v')$, whenever such a comparison is made in the verification algorithm steps described in Sect.~\ref{subsec:basicSearchProcess} and Sect.~\ref{subsec:inferredNodeDetection}. Since the neural networks learned using the \texttt{Caffe}~\cite{jia2014caffe} deep learning framework (which we employ for our experiments in this paper) tend not to degenerate in the node weights, this is sufficient for the experimental evaluation in this paper. Also, we did not observe any differences in the verification results between the SMT solver \texttt{Yices} \cite{DBLP:conf/cav/Dutertre14} on the SMT instances that we computed from the verification problems and the results computed by our tool.
The tool is available under the GPLv3 license and can be obtained from \texttt{https://github.com/progirep/planet} along with all scripts \& configuration files needed to learn the neural networks used in our experiments with the \texttt{Caffe} framework and to translate them to input files for our tool.

All computation times given in the following were obtained on a computer with an Intel Core i5-4200U 1.60\,GHz CPU and 8 GB of RAM running an x64 version of GNU/Linux. We do not report memory usage, as it was always $< 1$\,GB. All tools run with a single computation thread.

\subsection{Collision Avoidance}
\begin{figure}[tb]
{\centering\includegraphics[width=0.46\columnwidth]{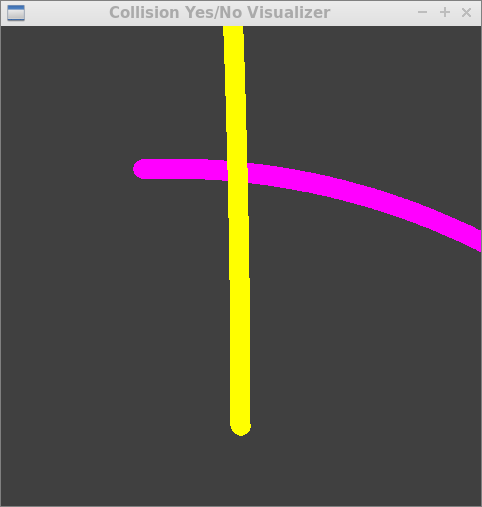}
$\quad$
\includegraphics[width=0.46\columnwidth]{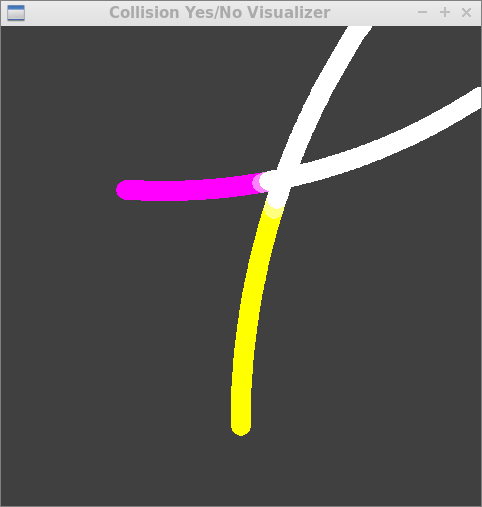}

}
\caption{Two pairs of vehicle trajectories, where the first one is non-colliding, and the second one is colliding. The lower vehicle starts roughly in north direction, whereas the other one starts roughly in east direction. The first trajectory is non-colliding as the two vehicles pass through the trajectory intersection point at different times.}
\label{fig:collisions}
\end{figure}

As a first example, we consider the problem of predicting collisions between two vehicles that follow curved paths at different speeds. We learned a neural network that processes tuples $(x,y,s,d,c_1,c_2)$ and classifies them into whether they represent a colliding or non-colliding case. In such a tuple,
\begin{itemize}
\item the $x$ and $y$ components represent the relative distances of the vehicles in their workspace in the X- and Y-dimensions,
\item the speed of the second vehicle is $s$,
\item the starting direction of the second vehicle is $d$, and
\item the rotation speed values of the two vehicles are $c_1$ and $c_2$.
\end{itemize}
The data is given in normalized (scaled) form to the neural network learner, so that all tuple components are between $0$ and $1$ (or between $-1$ and $1$ for $c_1$ and $c_2$). We wrote a tool that generates a few random tuples (within some intervals of possible values) along with  whether they represent a colliding or non-colliding case, as determined by simulation. The vehicles are circle-shaped, and we defined a safety margin and only consider tuples for which either the safety margins around the vehicles never overlap, or the vehicles themselves collide. So when only the safety margins overlap, this represents a ``don't care'' case for the learner. The tool also visualizes the cases, and we show two example traces in Figure~\ref{fig:collisions}. The tool ensures that the number of colliding cases and non-colliding ones are the same in the case list given to the neural network learner (by discarding tuples whenever needed). We generated 3000 tuples in total as input for \texttt{Planet}.

We defined a neural network architecture that consists of 40 linear nodes in the first layer, followed by a layer of MapPool nodes, each having 4 input edges, followed by a layer of 19 ReLU nodes, and 2 ReLU nodes for the output layer. Since \texttt{Caffe} employs randomization to initialize the node weights, the accuracy of the computed network is not constant. In 86 out of 100 tries, we were able to learn a network with an accuracy of 100\%, i.e., that classifies all example tuples correctly.

\mathchardef\mhyphen="2D

We want to find out the \emph{safety margin} around the tuples, i.e., the highest value of $\epsilon > 0$ such that for every tuple $(x,y,s,d,c_1,c_2)$ that is classified to $b \in \{\mathit{colliding},\mathit{notColliding}\}$, we have that all other tuples $(x \pm \epsilon,y \pm \epsilon,s \pm \epsilon, d\pm \epsilon, c_1 \pm \epsilon, c_2 \pm \epsilon)$ are classified to $b$ by the network as well. We perform this check for the first 100 tuples in the list, use \newterm{bisection search} to test this for $\epsilon \in [0,0.05]$, and abort the search process if $\epsilon$ has been determined with a precision of $0.002$.

\begin{figure}
\input{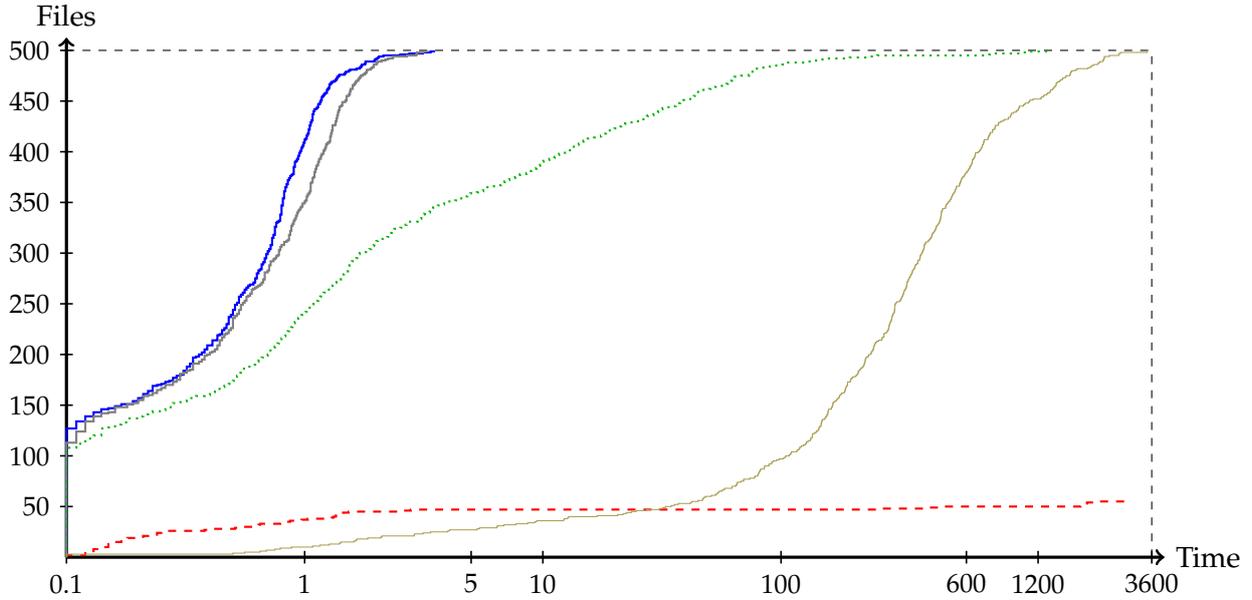}
\caption{Cactus plot of the solver time comparison for the 500 vehicle collision benchmarks. Time is given in seconds (on a $\log$-scale), and the lines, from bottom right to top left, represent \texttt{Gurobi} without linear approximation (dashed), \texttt{Yices} without linear approximation (solid), \texttt{Yices} with linear approximation (dotted), \texttt{Planet} (solid), and \texttt{Gurobi} with linear approximation (solid).}
\label{fig:cactusPlotCollision}
\end{figure}

We obtained 500 NN verification problem instances from this safety margin exploration process. Figure~\ref{fig:cactusPlotCollision} shows the distribution of the computation times of our tool on the problem instances, with a timeout of 1 hour. For comparison, we show the computation times of the SMT solver \texttt{Yices 2.5.2} and the (I)LP solver \texttt{Gurobi 7.02} on the problem instances. The SMT solver \texttt{z3} was observed to perform much worse than \texttt{Yices} on the verification problems, and is thus not shown. The choice of these comparison solvers was rooted in the fact that they performed best for verifying networks without MaxPool nodes in \cite{DBLP:journals/corr/KatzBDJK17}. We also give computation times for \texttt{Gurobi} and \texttt{Yices} after adding additional linear approximation constraints obtained with the approach in Section~\ref{subsec:linearApprox}. The computation times include the time to obtain them with our tool.

It can be observed that the computation times of \texttt{Gurobi} and \texttt{Yices} are too long for practical verification, except if the linear approximation constraints from our approach in this paper are added to the SMT and ILP instances to help the solvers. While \texttt{Yices} is then still slower than our approach, \texttt{Gurobi} actually becomes a bit faster in most cases, which is not surprising, given that it is a highly optimized commercial product that employs many sophisticated heuristics under-the-hood, whereas we use the less optimized \texttt{GLPK} linear programming framework. \texttt{Planet} spends most time on LP solving. It should be noted that the solver comparison is slightly skewed, as \texttt{Yices} employs arbitrary precision arithmetic whereas the other tools do not.

\subsection{MNIST Digit Recognition}
\label{subsec:digitRecognition}

As a second case study, we consider handwritten digit recognition. This is a classical problem in machine learning, and the MNIST dataset \cite{mnistlecun} is the most commonly used benchmark for comparing different machine learning approaches. The \texttt{Caffe} framework comes with some example architectures, and we use a simplified version of Caffe's version of the \texttt{lenet} network \cite{726791} for our experiments. The \texttt{Caffe} version differs from the original network in that is has piecewise linear node activation functions.

\begin{figure}[tb]
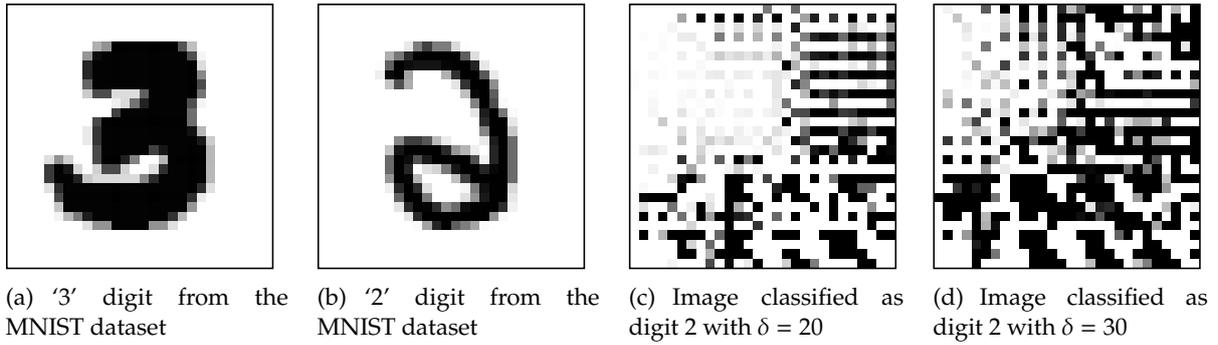

\begin{center}
\subfigure[`3' digit from the MNIST dataset]{\input{digit1} \label{fig:modelThree}} %485
$\ \ $
\subfigure[`2' digit from the MNIST dataset]{\input{digit2} \label{fig:hardlyIdentifiable2}} %
$\ \ $
\subfigure[Image classified as digit 2 with $\delta=20$]{\input{20er}\label{fig:20er}} %
$\ \ $
\subfigure[Image classified as digit 2 with $\delta=30$]{\input{30er}\label{fig:30er}} % \\
\end{center}

\caption{Example digit images from Section~\ref{subsec:digitRecognition}}
\label{fig:images}
\end{figure}

Figure~\ref{fig:images} (a)-(b) shows some example digits from the MNIST dataset. All images are in gray-scale and have $28 \times 28$ pixels. Our simplified network uses the following layers:
\begin{itemize}
\item One input layer with $28 \times 28$ nodes,
\item one convolutional network layer with $3 \times 13 \times 13$ nodes, where every node has 16 incoming edges,
\item one pooling layer with $3 \times 4 \times 4$ nodes, where each node has 16 incoming edges,
\item one ReLU layer with 8 nodes, and
\item one ReLU output layer with 10 nodes
\end{itemize}
The ReLU layers are fully connected. Overall, the network has 1341 nodes, the search space for the node phases is of size $16^{3 \cdot 4 \cdot 4} \cdot 2^8 \cdot 2^{10} = 2^{162}$, and the network has %3*16*13*13+3*4*4*(16+8)+8*10=
9344 edges.

We used this architecture to learn a network from the 100000 training images of the dataset, and the resulting network has an accuracy of 95.05\% on a separate testing dataset. Note that an accuracy of 100\% cannot be expected from any machine learning technique, as the dataset also contains digits that are even hardly identifiable for humans (as shown in Figure~\ref{fig:hardlyIdentifiable2}).

We performed a few tests with the resulting network. First we wanted to see an input image that is classified strongly as a $2$. More formally, we wanted to obtain an input image $(x_{1,1}, \ldots, x_{28,28})$ for which the network outputs a vector $(y_0, \ldots, y_9)$ for which $y_2 \geq y_i + \delta$ for all $i \in \{0,1,3,4,5,6,7,8,9\}$ for a large value of $\delta$. We found that for values of $\delta=20$ and $\delta=30$, such images can be found in 4 minutes 25 seconds and 32 minutes 35 seconds, respectively. The two images are shown in Figure~\ref{fig:20er} and Figure~\ref{fig:30er}. For $\delta = 50$, no such image can be found (4 minutes 41 seconds of computation time), but for $\delta = 35$, \texttt{Planet} times out after 4 hours. \texttt{Gurobi} (with the added linear approximation constraints) could not find a solution in this time frame, either.

Then, we are interested in how much noise can be added to images before they are not categorized correctly anymore. We start with the digit given in Figure~\ref{fig:modelThree}, which is correctly categorized by the learned network as digit 3. We ask whether there is another image that is categorized as a 4, but for which each pixel has values that are within an absolute range of $\pm 8 \%$  of color intensity of the original image's pixels, where we keep the pixels the same that are at most three pixels away from the boundaries. To determine that this is not the case, \texttt{planet} requires 1 minutes 46.8 seconds. For a range of  $\pm 0.12$, \texttt{planet} times out after four hours. The output of \texttt{planet} shows that long conflict clauses are learned in the process, which suggests that we applied it to a difficult verification problem.

We then considered an error model that captures noise that is likely to occur in practice (e.g., due to stains on scanned paper). It excludes sharp noise edges such as the ones in Figure~\ref{fig:30er}. Instead of restricting the amplitude of noise, we restrict the noise value differences in adjacent pixels to be $\leq 0.05$ (i.e., 5\% of color density). This constraint essentially states that the noise must pass through a linearized low-pass filter unmodified. We still exclude the pixels from the image boundaries from being modified. 
Our tool concludes in 9 minutes 2.4 seconds that the network never misclassifies the image from Figure~\ref{fig:modelThree} as a $4$ under this noise model. Since the model allows many pixels to have large deviations, we can see that including a linear noise model can improve the computation time of \texttt{planet}.

\section{Conclusion}

In this paper, we presented a new approach for the verification of feed-forward neural networks with piece-wise linear activation functions. Our main idea was to generate a linear approximation of the overall network behavior that can be added to SMT or ILP instances which encode neural network verification problems, and to use the approximation in a specialized approach that features multiple additional techniques geared towards neural network verification, which are grouped around a SAT solver for choosing the node phases in the network. We considered two case studies from different application domains. The approach allows arbitrary convex verification conditions, and we used them to define a noise model for testing the robustness of a network for recognizing handwritten digits.

We made the approach presented in this paper available as open-source software in the hope that it fosters the co-development of neural network verification tools and neural network architectures that are easier to verify. While our approach is limited to network types in which all components have piece-wise linear activation functions, they are often the only ones used in modern network architectures anyway. But even if more advanced activation functions such as \newterm{exponential linear units} \cite{DBLP:journals/corr/ClevertUH15} shall be used in learning, they can still be applied to learn an initial model, which is then linearly approximated with ReLU nodes and fine-tuned by an additional learning process. The final model is then easier to verify. Such a modification of the network architecture during the learning process is not commonly applied in the artificial intelligence community yet, but while verification becomes more practical, this may change in the future.

Despite the improvement in neural network verification performance reported in this paper, there is still a lot to be done on the verification side: we currently do not employ specialized heuristics for node phase branching selection, and while our approach increases the scalablity of neural network verification substantially, we observed it to still be quite fragile and prone to timeouts for difficult verification properties (as we saw in the MNIST example).
Also, we had to simplify the LeNet architecture for digit recognition in our experiments, as the original net is so large that even obtaining a lower bound for a single variable in the network (which we do for all network nodes before starting the actual solution process as explained in Section~\ref{subsec:linearApprox}) takes more than 30 minutes otherwise, even though this only means solving a single linear program. While the approach by Huang et al.~\cite{DBLP:journals/corr/HuangKWW16} does not suffer from this limitation, it cannot handle general verification properties, which we believe to be important. We plan to work on tackling the network size limitation of the approach presented in this paper in the future.

\section*{Acknowledgements}
This work was partially funded by the Institutional Strategy of the University of Bremen, funded by the German Excellence Initiative.

\bibliographystyle{alpha}
\bibliography{bib}

\newcommand{\etalchar}[1]{$^{#1}$}
\begin{thebibliography}{HKWW17}

\bibitem[BIL{\etalchar{+}}16]{DBLP:conf/nips/BastaniILVNC16}
Osbert Bastani, Yani Ioannou, Leonidas Lampropoulos, Dimitrios Vytiniotis,
  Aditya~V. Nori, and Antonio Criminisi.
\newblock Measuring neural net robustness with constraints.
\newblock In {\em Annual Conference on Neural Information Processing Systems
  (NIPS)}, pages 2613--2621, 2016.

\bibitem[CD91]{DBLP:journals/informs/ChinneckD91}
John~W. Chinneck and Erik~W. Dravnieks.
\newblock Locating minimal infeasible constraint sets in linear programs.
\newblock {\em {INFORMS} Journal on Computing}, 3(2):157--168, 1991.

\bibitem[CUH15]{DBLP:journals/corr/ClevertUH15}
Djork{-}Arn{\'{e}} Clevert, Thomas Unterthiner, and Sepp Hochreiter.
\newblock Fast and accurate deep network learning by exponential linear units
  ({ELUs}).
\newblock {\em arXiv/CoRR}, 1511.07289, 2015.

\bibitem[DdM06]{DBLP:conf/cav/DutertreM06}
Bruno Dutertre and Leonardo~Mendon{\c{c}}a de~Moura.
\newblock A fast linear-arithmetic solver for {DPLL(T)}.
\newblock In {\em 18th International Conference on Computer Aided Verification
  (CAV)}, pages 81--94, 2006.

\bibitem[Dut14]{DBLP:conf/cav/Dutertre14}
Bruno Dutertre.
\newblock Yices 2.2.
\newblock In {\em 26th International Conference on Computer Aided Verification
  (CAV)}, pages 737--744. Springer, 2014.

\bibitem[ES03]{DBLP:conf/sat/EenS03}
Niklas E{\'{e}}n and Niklas S{\"{o}}rensson.
\newblock An extensible {SAT}-solver.
\newblock In {\em 6th International Conference on Theory and Applications of
  Satisfiability Testing, ({SAT}). Selected Revised Papers}, pages 502--518,
  2003.

\bibitem[FM09]{FM09HBSAT}
John Franco and John Martin.
\newblock {\em A History of Satisfiability}, volume 185 of {\em Frontiers in
  Artificial Intelligence and Applications}, chapter~1, pages 3--74.
\newblock IOS Press, February 2009.

\bibitem[HKWW17]{DBLP:journals/corr/HuangKWW16}
Xiaowei Huang, Marta Kwiatkowska, Sen Wang, and Min Wu.
\newblock Safety verification of deep neural networks.
\newblock In {\em 29th International Conference on Computer Aided Verification
  (CAV)}. Springer, 2017.

\bibitem[JSD{\etalchar{+}}14]{jia2014caffe}
Yangqing Jia, Evan Shelhamer, Jeff Donahue, Sergey Karayev, Jonathan Long, Ross
  Girshick, Sergio Guadarrama, and Trevor Darrell.
\newblock Caffe: Convolutional architecture for fast feature embedding.
\newblock {\em arXiv/CoRR}, 1408.5093, 2014.

\bibitem[KBD{\etalchar{+}}17]{DBLP:journals/corr/KatzBDJK17}
Guy Katz, Clark~W. Barrett, David~L. Dill, Kyle Julian, and Mykel~J.
  Kochenderfer.
\newblock Reluplex: An efficient {SMT} solver for verifying deep neural
  networks.
\newblock In {\em 29th International Conference on Computer Aided Verification
  (CAV)}. Springer, 2017.

\bibitem[KS08]{Kroening2008}
Daniel Kroening and Ofer Strichman.
\newblock {\em Decision Procedures -- An Algorithmic Point of View}.
\newblock Springer, 2008.

\bibitem[LBBH98]{726791}
Y.~Lecun, L.~Bottou, Y.~Bengio, and P.~Haffner.
\newblock Gradient-based learning applied to document recognition.
\newblock {\em Proceedings of the IEEE}, 86(11):2278--2324, 1998.

\bibitem[LC09]{mnistlecun}
Yann Lecun and Corinna Cortes.
\newblock {The MNIST database of handwritten digits}, 2009.

\bibitem[PT10]{DBLP:conf/cav/PulinaT10}
Luca Pulina and Armando Tacchella.
\newblock An abstraction-refinement approach to verification of artificial
  neural networks.
\newblock In {\em 22nd International Conference on Computer Aided Verification
  ({CAV})}, pages 243--257, 2010.

\bibitem[PT12]{DBLP:journals/aicom/PulinaT12}
Luca Pulina and Armando Tacchella.
\newblock Challenging {SMT} solvers to verify neural networks.
\newblock {\em {AI} Commun.}, 25(2):117--135, 2012.

\bibitem[Sch15]{DBLP:journals/nn/Schmidhuber15}
J{\"{u}}rgen Schmidhuber.
\newblock Deep learning in neural networks: An overview.
\newblock {\em Neural Networks}, 61:85--117, 2015.

\bibitem[SNM{\etalchar{+}}16]{DBLP:conf/fmcad/ScheiblerNMFTBF16}
Karsten Scheibler, Felix Neubauer, Ahmed Mahdi, Martin Fr{\"{a}}nzle, Tino
  Teige, Tom Bienm{\"{u}}ller, Detlef Fehrer, and Bernd Becker.
\newblock Accurate {ICP}-based floating-point reasoning.
\newblock In {\em Formal Methods in Computer-Aided Design (FMCAD)}, pages
  177--184, 2016.

\bibitem[SWWB15]{DBLP:conf/mbmv/ScheiblerWWB15}
Karsten Scheibler, Leonore Winterer, Ralf Wimmer, and Bernd Becker.
\newblock Towards verification of artificial neural networks.
\newblock In {\em MBMV Workshop 2015, Chemnitz, Germany}, pages 30--40, 2015.

\bibitem[WK15]{Wagner2015}
Michael Wagner and Philip Koopman.
\newblock A philosophy for developing trust in self-driving cars.
\newblock In {\em Road Vehicle Automation 2}, pages 163--171. Springer
  International Publishing, 2015.

\bibitem[YYS{\etalchar{+}}15]{DBLP:conf/bmvc/YuYSXH15}
Qian Yu, Yongxin Yang, Yi{-}Zhe Song, Tao Xiang, and Timothy~M. Hospedales.
\newblock Sketch-a-net that beats humans.
\newblock In {\em British Machine Vision Conference (BMVC)}, pages 7.1--7.12,
  2015.

\end{thebibliography}

\end{document}